\documentclass[letterpaper]{article} 
\usepackage{aaai2026}  
\usepackage{times}  
\usepackage{helvet}  
\usepackage{courier}  
\usepackage[hyphens]{url}  
\usepackage{graphicx} 
\usepackage{natbib}  

\usepackage{caption} 
\usepackage{algorithm}
\usepackage{algorithmic}
\usepackage{pgfplots}
\pgfplotsset{compat=1.18}

\usepackage{amsmath, amsfonts, amssymb, booktabs}

\usepackage{newfloat}
\usepackage{listings}
\DeclareCaptionStyle{ruled}{labelfont=normalfont,labelsep=colon,strut=off} 
\floatstyle{ruled}
\newfloat{listing}{tb}{lst}{}
\floatname{listing}{Listing}
\title{Boosting ASR Robustness via Test-Time Reinforcement Learning\\ with Audio-Text Semantic Rewards}
\author{
    Linghan Fang\textsuperscript{\rm 1,2},
    Tianxin Xie\textsuperscript{\rm 1},
   Li Liu\textsuperscript{\rm 1}\thanks{Corresponding author.}
}

\affiliations{
    \textsuperscript{\rm 1}The Hong Kong University of Science and Technology (Guangzhou), Guangzhou, China\\
    \textsuperscript{\rm 2}Technical University of Munich, Munich, Germany\\
    linghan.fang@tum.de,
    txie151@connect.hkust-gz.edu.cn,
    avrillliu@hkust-gz.edu.cn
}

\usepackage{bibentry}

\begin{document}

\maketitle

\begin{abstract}

Recently, Automatic Speech Recognition (ASR) systems (e.g., Whisper) have achieved remarkable accuracy improvements but remain highly sensitive to real-world unseen data (data with large distribution shifts), including noisy environments and diverse accents. To address this issue, test-time adaptation (TTA) has shown great potential in improving the model adaptability at inference time without ground-truth labels, and existing TTA methods often rely on pseudo-labeling or entropy minimization. However, by treating model confidence as a learning signal, these methods may reinforce high-confidence errors, leading to confirmation bias that undermines adaptation.
To overcome these limitations, we present \textbf{ASR-TRA}, a novel \textbf{T}est-time \textbf{R}einforcement \textbf{A}daptation framework inspired by causal intervention. More precisely, our method introduces a learnable decoder prompt and utilizes temperature-controlled stochastic decoding to generate diverse transcription candidates. These are scored by a reward model that measures audio-text semantic alignment, and the resulting feedback is used to update both model and prompt parameters via reinforcement learning.
Comprehensive experiments on LibriSpeech with synthetic noise and L2 Arctic accented English datasets demonstrate that our method achieves higher accuracy while maintaining lower latency than existing TTA baselines. Ablation studies further confirm the effectiveness of combining audio and language-based rewards, highlighting our method's enhanced stability and interpretability. Overall, our approach provides a practical and robust solution for deploying ASR systems in challenging real-world conditions.
\end{abstract}

\begin{links}
    \link{Code}{https://github.com/fangcq/ASR-TRA}
\end{links}

\section{Introduction}
Recent advances in automatic speech recognition (ASR) have been driven by breakthroughs in self‑supervised learning and large‑scale weakly supervised training
~\cite{baevski2020wav2vec,hsu2021hubert,schneider2019wav2vec},
which enables models to learn rich acoustic and linguistic representations from vast amounts of unlabeled or loosely labeled speech data. Leveraging these techniques, models such as wav2vec 2.0~\cite{baevski2020wav2vec} and Whisper~\cite{radford2023whisper} have significantly improved transcription accuracy and generalization in various domains.

Despite these improvements, deploying ASR in real‑world applications such as edge devices, online streaming, and resource‑constrained small‑model scenarios remains challenging.
In practice, lightweight ASR systems often face severe out‑of‑distribution (OOD) conditions~\cite{hendrycks2019ood}, including background noise, heavy accents, and regional dialects, which are under‑represented in training data and lead to significant domain shifts~\cite{ben2007domain}.
Traditional methods to improve robustness typically rely on offline retraining or supervised domain adaptation, such as multi‑condition training or noise augmentation~\cite{ko2015audio,li2020multi}, but these approaches are infeasible during test time because labeled data is unavailable.

\begin{figure}[t]
  \centering
  \includegraphics[width=\linewidth]{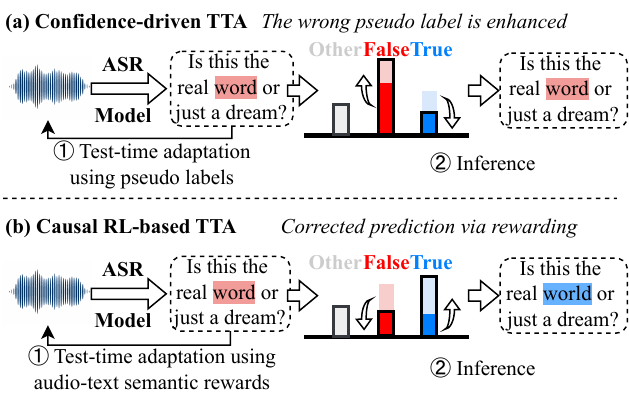}
  \caption{
    \textbf{Overview of test-time adaptation strategies under noisy conditions.}
    (a) Confidence-driven adaptation reinforces incorrect but high-probability predictions (e.g., \emph{word}), leading to persistent errors.
    (b) Our reward-guided adaptation \textbf{ASR-TRA} favors semantically accurate alternatives (e.g., \emph{world}) through reinforcement learning, even when their initial confidence is low.
    }
  \label{fig:demo}
\end{figure}


\begin{figure}[t]
    \centering
    \includegraphics[width=0.85\linewidth]{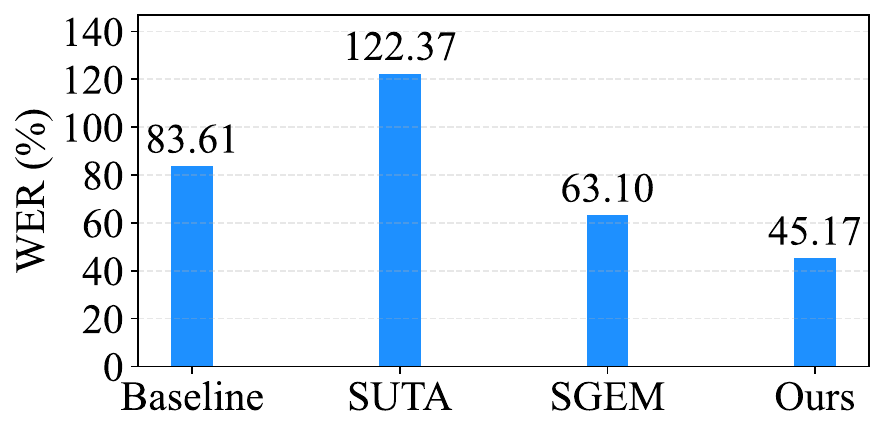}
    \caption{We evaluate WER on the top-100 high-confidence samples from LibriSpeech test-other corrupted with Gaussian noise. While baseline performance from Whisper-Tiny suffers under noise, heuristic methods like SUTA further degrade due to overconfidence. In contrast, ours achieves lower WER. (Lower is better.)}
    \label{fig:confident-subset}
\end{figure}

In the literature, test-time adaptation (TTA) has emerged as a promising approach to enhance ASR robustness without requiring additional training or labeled supervision~\cite{wang2021tent, lin2022suta, kim2023sgem}. By enabling models to adapt dynamically to test-time inputs, TTA helps mitigate the mismatch between training and deployment conditions.
However, current TTA methods have the following two main challenges, when directly apply to ASR systems (see Fig.~\ref{fig:demo}). \textbf{Challenge 1:} Although Whisper is a powerful encoder-decoder model, it lacks a Whisper‑specific adaptation mechanism~\cite{radford2023whisper}. \textbf{Challenge 2:} Many existing approaches rely on heuristic optimization strategies and heavily use self‑generated pseudo‑labels~\cite{wang2021tent,lin2022suta,lee2013pseudo,sohn2020fixmatch}, which can be slow, unstable in acoustically diverse environments, and prone to compounding errors when the model is uncertain. As illustrated in Figure~\ref{fig:confident-subset}, confidence‑based methods such as SUTA~\cite{lin2022suta} can perform even worse on high‑confidence samples under noise, highlighting the misalignment between confidence and true accuracy that motivates our approach (see the Experiments section for details).

To address these challenges, we present a novel method \textbf{ASR-TRA} (\textbf{ASR} with \textbf{T}est-time \textbf{R}einforcement \textbf{A}daptation), which adopts a reinforcement learning (RL)~\cite{williams1992reinforce,ziegler2019rlhf} perspective and frames TTA as a reward‑driven decision process under uncertainty.
ASR-TRA addresses these limitations based on two key ideas.
\textbf{Idea 1:} To leverage the structure of encoder-decoder architectures such as Whisper, we introduce a dedicated adaptation pathway through a learnable decoder prompt, enabling efficient and low-overhead test-time optimization tailored to ASR.
\textbf{Idea 2:} Instead of relying on fragile heuristics or self-confirming pseudo-labels, we introduce an external semantic reward to guide adaptation more robustly. This avoids compounding errors and allows for more stable updates without ground-truth supervision.


\begin{figure}[t]
  \centering
  \includegraphics[width=0.8\linewidth]{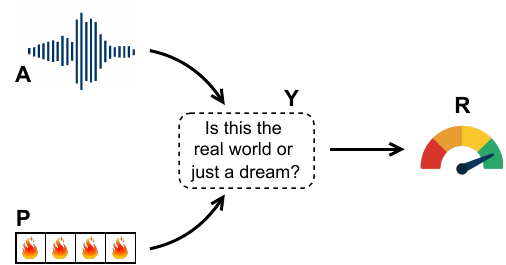}
  \caption{Structural Causal Model (SCM) schematic.
           Nodes \(A\), \(P\), \(Y\), and \(R\) denote audio features, learnable prompt,
           transcription output, and reward, respectively, with causal flow
           \(A,P \!\rightarrow\! Y \!\rightarrow\! R\)%
           ~\cite{pearl2009causality}.}
  \label{fig:SCM}
\end{figure}


To realize these two key ideas, ASR-TRA begins with a structured modeling of the adaptation process using a Structural Causal Model (SCM)~\cite{pearl2009causality}, which bridges the motivation with concrete implementation. The SCM comprises four key variables: the encoded audio features \( A \), the learnable decoder prompt \( P \), the generated transcription \( Y \), and the reward \( R \). The directed edges capture the causal dependencies among these components (see Figure~\ref{fig:SCM}). 

This model not only clarifies how adaptation unfolds in our framework, but also guides the design of three concrete components that operationalize reward-driven learning and prompt-based adaptation at test time:
1) \textbf{Prompt injection (for Idea1):} Insert a learnable vector \(P\) at the beginning of the decoder input sequence, so that \(P\) is processed alongside the token embeddings and directly influences the decoding process as a causal intervention.
2) \textbf{Candidate generation \& evaluation (for Idea2):} Adjust the sampling temperature to produce multiple diverse transcription hypotheses. Each candidate is then evaluated using CLAP (Contrastive Language–Audio Pretraining)~\cite{elizalde2023clap}, which can compute an audio–text similarity score as a sequence‑level reward to guide adaptation.
3) \textbf{Parameter update:} Apply a policy‑gradient RL algorithm to backpropagate CLAP rewards and jointly update both the prompt parameters \(P\) and the model weights, steering future outputs toward higher reward.

In summary, the proposed ASR-TRA approach offers three main contributions:
\begin{itemize}
    \item We formulate TTA as an RL process guided by the audio‑text reward model CLAP, which mitigates error accumulation from heuristic pseudo‑label or confidence‑based methods.
    \item We design a Whisper‑specific Structural Causal Model (SCM) where a learnable decoder prompt modulates the decoding process.
    Combined with policy‑gradient updates and the CLAP reward, this framework enables a principled and lightweight TTA approach for ASR.
    \item Experiments with noisy and accented speech benchmarks show that it consistently outperforms previous TTA methods in both speed and recognition accuracy.
\end{itemize}

\section{Related Work}

\paragraph{Automatic Speech Recognition.} 
Recent progress in automatic speech recognition has been largely driven by the development of large-scale neural network models trained on diverse audio-text corpora. Wav2vec 2.0~\cite{baevski2020wav2vec} leverages self-supervised contrastive pretraining to learn robust speech representations from unlabeled audio. Whisper~\cite{radford2023whisper}, built on a transformer encoder-decoder architecture, achieves strong multilingual transcription by training on 680k hours of weakly labeled data. Despite their impressive performance, these models remain sensitive to distribution shifts, such as environmental noise, speaker accents, or spontaneous speech, underscoring the need for adaptive mechanisms ~\cite{hendrycks2019ood,ben2007domain} that can generalize well to real-world conditions.

\paragraph{Causality in Deep Learning.} 
Causal reasoning offers a principled framework for improving model robustness and interpretability. Classical formulations~\cite{pearl2009causality} define interventions as mechanisms for actively altering model variables to study downstream effects. Recent work has explored causal representation learning~\cite{scholkopf2021toward,mitrovic2021representation}, which aims to disentangle generative factors of variation for greater robustness under domain shift. In deep learning, causal interventions can be realized through architectural modifications or controlled sampling schemes and have been shown to help models adapt more reliably to distributional changes.

\paragraph{Test-Time Adaptation in ASR.} 
TTA enables models to adjust during inference using only unlabeled test data. Among early approaches, Tent~\cite{wang2021tent} proposed an \emph{episodic} TTA framework that adapts the model to each test sample by minimizing entropy while freezing most parameters. This inspired a range of episodic TTA methods that aim for rapid and lightweight domain adaptation. In the ASR domain, SUTA~\cite{lin2022suta} adapts Whisper-style models by generating pseudo-labels from high-confidence predictions, while SGEM~\cite{kim2023sgem} minimizes sequence-level entropy to regularize the output distribution. These methods, however, rely heavily on model-internal signals such as entropy or confidence, which can become unreliable under severe distribution shift. Furthermore, they lack a mechanism to incorporate feedback external to the model’s prior beliefs.

\paragraph{Reinforcement Learning for TTA.}
RL offers a general framework for optimizing decisions based on delayed feedback~\cite{sutton2018reinforcement}, and has been widely applied to robotics~\cite{levine2016end}, dialogue~\cite{li2016deep}, and language model alignment~\cite{ouyang2022training,ziegler2019rlhf}. Recently, RL has also been explored in test-time adaptation. For instance, RLCF~\cite{Zhao2024TestTimeCLIP} learns reward-driven policies for classifier adaptation. More recent efforts, such as BiTTA~\cite{lee2025bitta}, leverage binary feedback (correct/incorrect) to guide episodic adaptation. While promising, these methods have been mostly applied to classification or navigation tasks, where the action space and feedback signals are comparatively straightforward. In contrast, their application to structured sequence modeling problems such as ASR remains limited.

\paragraph{Positioning of ASR-TRA.}
While prior TTA methods in ASR, such as SUTA and SGEM, rely on internal heuristics like confidence and entropy, they often lack robustness under distribution shift and may suffer from pseudo-label feedback loops. Recent RL-based approaches such as BiTTA demonstrate the utility of reward signals, but are designed for classification or navigation tasks and have not been adapted for structured sequence modeling in ASR. ASR-TRA bridges this gap by combining causal interventions with external reward-guided adaptation in a Whisper-based framework. Specifically, we introduce learnable decoder prompts as causal variables and optimize them using semantic feedback~\cite{elizalde2023clap}, enabling robust, interpretable, and efficient adaptation without relying on internal certainty estimates.

\section{Method}

We propose a TTA framework, ASR-TRA, for Whisper that integrates causal reasoning with reinforcement learning. The key insight is to treat prompt injection in the Whisper decoder as a causal intervention and the decoding process as a generator of counterfactual hypotheses. We optimize the model at inference time by rewarding generations that better align with CLAP's predictions, without relying on ground-truth transcripts. This enables dynamic and label-free adaptation to unseen acoustic conditions.

\subsection{Whisper Architecture}

The input speech signal is first preprocessed and converted into a log Mel‑spectrogram $s \in \mathbb{R}^{F \times T}$ with $F$ frequency bins and $T$ time frames \citep{logan2000mel,mcfee2015librosa}.  
Whisper is a Transformer‑based encoder–decoder model \citep{vaswani2017attention,radford2023whisper} that follows an autoregressive sequence‑to‑sequence formulation widely used in ASR \citep{chan2016listen}.  
The audio encoder $\mathrm{Enc}(\cdot)$ produces a hidden representation:
\begin{equation}
    h = \mathrm{Enc}(s),
\end{equation}
and the text decoder generates the output sequence $\hat{y} = (y_1, y_2, \ldots, y_N)$ by modeling the conditional token distribution over output probabilities:
\begin{equation}
    P(y_t \mid y_{<t}, h) = \mathrm{Dec}(y_{<t}, h),
\end{equation}
where $y_{<t}$ denotes previously emitted tokens.  
In practice, the decoding process can be performed using greedy selection, beam search, or stochastic sampling regulated by temperature scaling \citep{holtzman2020curious}.

\begin{figure*}
    \centering
    \includegraphics[width=\linewidth]{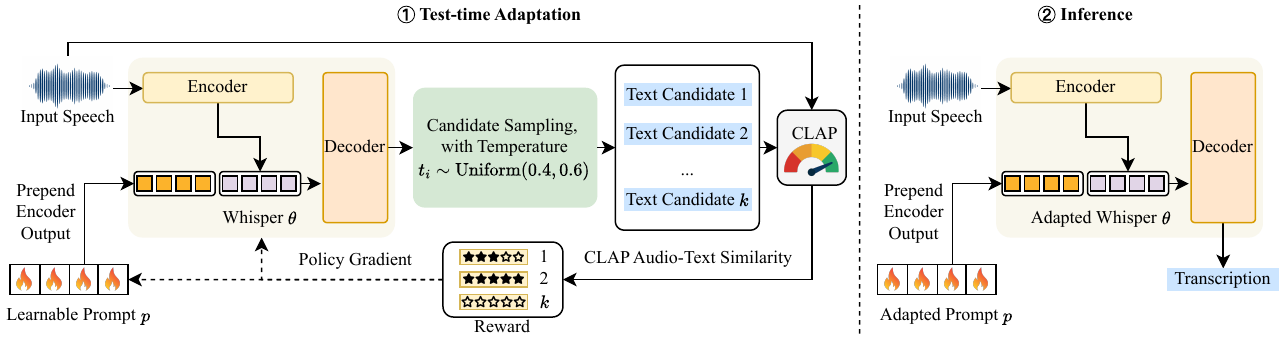}
    \caption{Test‑time self‑adaptation for Whisper. A baseline transcript is decoded from the input Mel‑spectrogram; soft‑prompted variants are then sampled at varied temperatures, CLAP scores them all, and the aggregated rewards update the model, achieving on‑the‑fly correction without labels.}

    \label{fig:enter-label}
\end{figure*}

\subsection{Prompt Injection as Causal Intervention}

We introduce a learnable prompt vector $p \in \mathbb{R}^{L \times d}$, where $L$ denotes the number of prompt tokens and $d$ is the decoder embedding dimension. A soft prompt is prepended to the decoder’s input embeddings, directly concatenated before the embedding of the \texttt{\textless{}bos\textgreater{}} token. During generation, the decoder attends to both this prompt and the encoder output, allowing the prompt to guide each prediction instead of relying solely on autoregressive decoding from scratch.

Because the prompt is visible to the decoder’s attention at every step, it can directly shape the hidden states and thus influence all subsequent token predictions \citep{li2021prefix,lester2021prompt}.
  
Crucially, we treat $p$ not merely as an additional condition but as a \emph{causal intervention} on the generation process. Formally, the decoder prediction can be written as
\begin{equation}
    y_t = \mathrm{Dec}(h, y_{<t}, \mathrm{do}(p)),
\end{equation}
where $\mathrm{do}(p)$ follows Pearl's do‑calculus \citep{pearl2009causality} and indicates that $p$ is set externally rather than inferred from the observed input $s$. This intervention perturbs the internal generation dynamics without modifying the acoustic input, enabling the model to explore alternative hypotheses under the same observation.  

\subsection{Counterfactual Sampling and Reward Evaluation}

To explore diverse output trajectories under a fixed input and prompt, we adopt stochastic decoding by sampling tokens with a temperature parameter $T > 0$ \citep{holtzman2020curious}: 
\begin{equation}
    P_T(y_t \mid y_{<t}, h, p) 
    \propto 
    \exp\left(\frac{\log P(y_t \mid y_{<t}, h, p)}{T}\right).
\end{equation}

A higher temperature $T$ flattens the token distribution and encourages diversity, while $T \to 0$ approaches greedy decoding.  
By sampling repeatedly, we obtain $K$ candidate transcriptions 
$\{\hat{y}^{(1)}, \ldots, \hat{y}^{(K)}\}$, each representing a \emph{counterfactual hypothesis}, i.e., a plausible alternative transcription trajectory conditioned on the same audio and prompt.

Each sampled hypothesis is then evaluated using a reward function 
$R(\hat{y}^{(i)}) \in \mathbb{R}$ that quantifies its semantic alignment with the input.  
In our implementation, we adopt the CLAP audio–text similarity model \citep{elizalde2023clap} as the primary reward, which computes the cosine
similarity between audio embeddings of input speech and text embeddings of generated transcriptions: 
\begin{equation}
    r^{(i)} = R(\hat{y}^{(i)}).
\end{equation}
We also use the scores from other pretrained language models (LM) \citep{ziegler2019rlhf,ouyang2022training} as a supplementary signal in our ablation study (Table~\ref{tab:ablation-wer-latency}).

This scalar feedback serves as a proxy for transcription quality and enables label‑free optimization.  
The resulting rewards are aggregated and used to update both the prompt vector $p$ and model parameters via a policy‑gradient objective \citep{williams1992reinforce,sutton2018reinforcement}, thereby biasing the model towards generations that achieve higher reward under the same conditions.

\subsection{Optimization via Reinforcement Learning}

To adapt the prompt vector $p$  and model parameters online, we formulate a reinforcement learning objective that explicitly encourages generations with higher semantic quality as measured by the reward function.  
We treat the prompt‑conditioned Whisper as a stochastic policy $\pi_p(\hat{y})$ over output sequences $\hat{y}$ and aim to maximize the expected reward:
\begin{equation}
    \mathcal{J}(p) = \mathbb{E}_{\hat{y} \sim \pi_p}[R(\hat{y})] .
\end{equation}

This setting naturally aligns with policy‑gradient methods \citep{sutton2018reinforcement}.  
We apply the classic REINFORCE algorithm \citep{williams1992reinforce} to estimate gradients of $\mathcal{J}(p)$.  
For a batch of $N$ sampled hypotheses 
$\{\hat{y}^{(i)}\}_{i=1}^N$ 
with corresponding rewards $r^{(i)}$ and log‑probabilities 
$\log P(\hat{y}^{(i)})$, we introduce a baseline to reduce gradient variance.  
The baseline is simply the mean reward across the batch:
\begin{equation}
    \bar{r} = \frac{1}{N} \sum_{i=1}^N r^{(i)}.
\end{equation}
Using this baseline, the gradient of the RL objective with respect to the learnable prompt $p$ and Whisper's parameters $\theta$ can be estimated as follows:
\begin{equation}
    \nabla_{\theta,p} \mathcal{L} = 
    -\frac{1}{N} \sum_{i=1}^N 
    \nabla_{\theta,p} \log P(\hat{y}^{(i)}) \cdot (r^{(i)} - \bar{r}),
\end{equation}
which increases the likelihood of high‑reward generations while penalizing low‑reward alternatives.  

During test time, we perform optimization independently for each input sample, using a small number of reward‑evaluated candidates to update the prompt and model parameters.
Once the adaptation and prediction for the current sample is completed, the model parameters are restored to their original state before processing the next sample, ensuring that updates do not accumulate across the test set.

\subsection{Algorithm Overview}
We summarize our TTA method ASR-TRA in \textbf{Algorithm~\ref{alg:tta-whisper}}, which shows how we integrate prompt intervention, counterfactual sampling, and reinforcement learning updates into a unified online ``adaptation-and-prediction'' loop.
\begin{algorithm}[H]
\caption{\textbf{ASR-TRA}}
\label{alg:tta-whisper}
\begin{algorithmic}[1]
\REQUIRE Input Mel-spectrogram $s$, Whisper model $\theta$, CLAP reward model, number of samples $n$, prompt embedding $p$
\STATE Initialize Whisper model $\theta$, set temperature $t=0$
\STATE Generate deterministic output: $y_0 \leftarrow \text{Whisper}(s; \theta, t=0)$
\STATE Compute reward: $r_0 \leftarrow \text{CLAP}(y_0, s)$
\STATE Insert random prompt $p$ into Whisper decoder
\FOR{$i = 1,\dots,n$}
    \STATE Sample temperature $t_i \sim \text{Uniform}(0.4, 0.6)$
    \STATE Generate output with intervention:\\
    $y_i \leftarrow \text{Whisper}(s; \theta, \mathrm{do}(p), t_i)$
    \STATE Compute reward: $r_i \leftarrow \text{CLAP}(y_i, s)$
\ENDFOR
\STATE Compute advantages: $\text{adv}_i \leftarrow r_i - \text{mean}(\{r_j\}_{j=0}^n)$
\STATE Compute policy gradient loss: \\
$\mathcal{L} \leftarrow -\sum_i \text{adv}_i \cdot \log P_\theta(y_i|s,p)$
\STATE Update model parameters and prompt: \\
$\theta \leftarrow \theta - \eta_1 \nabla_\theta \mathcal{L}, \; p \leftarrow p - \eta_2 \nabla_p \mathcal{L}$
\STATE Generate adapted output: $y \leftarrow \text{Whisper}(s; \theta, p)$
\STATE Restore Whisper model $\theta$ and prompt $p$ to original state
\RETURN Adapted transcription $y$
\end{algorithmic}
\end{algorithm}


\begin{table*}[t]
\small
\centering

\begin{tabular}{c|c|cc|cc|cc}
\toprule
 & Baseline & \multicolumn{2}{c|}{+ SUTA} & \multicolumn{2}{c|}{+ SGEM} & \multicolumn{2}{c}{+ ASR-TRA (Ours)} \\
Dataset & WER (\%) $\downarrow$ & WER (\%) $\downarrow$ & Latency (s) & WER (\%) $\downarrow$  & Latency (s) & WER (\%) $\downarrow$ & Latency (s) \\
\midrule
AC   & 26.54 & 24.02 & 1.683 & 23.76 & 0.730 & \textbf{22.39} & \textbf{0.678} \\
AA   & 40.22 & 37.69 & 1.770 & 36.66 & 0.759 & \textbf{35.66} & \textbf{0.670} \\
BA   & 36.85 & 40.12 & 1.643 & 33.92 & 0.710 & \textbf{31.39} & \textbf{0.679} \\
CM   & 39.40 & 43.52 & 1.583 & 35.02 & 0.691 & \textbf{34.72} & \textbf{0.631} \\
MU   & 29.98 & 34.12 & 1.645 & 34.40 & 0.705 & \textbf{25.88} & \textbf{0.584} \\
NB   & 40.77 & \textbf{33.75} & 1.847 & 36.13 & \textbf{0.756} & 35.81 & 0.833 \\
SD   & 23.12 & 22.91 & 1.656 & \textbf{19.50} & 0.909 & 22.34 & \textbf{0.801} \\
TP   & 24.81 & 22.05 & 1.699 & 22.40 & 0.941 & \textbf{20.91} & \textbf{0.886} \\
\midrule
Mean & 32.71 & 32.27 & 1.690 & 30.22 & 0.775 & \textbf{28.64} & \textbf{0.720} \\
\bottomrule
\end{tabular}
\caption{WER (\%) and inference latency (seconds) on Whisper-Tiny under eight noise conditions from MS-SNSD (SNR = 10~dB). Best values per row are in \textbf{bold}.}
\label{tab:main-results}
\end{table*}

\begin{table}
\setlength{\tabcolsep}{2pt}
\centering
\begin{tabular}{lcccc}
\toprule
Setting & Baseline & SUTA & SGEM & ASR-TRA (Ours) \\
\midrule
Arabic      & 32.74 & 35.94 & 35.19 & \textbf{22.92} \\
Mandarin    & 28.03 & 27.68 & 28.03 & \textbf{25.14} \\
Hindi       & 14.62 & 14.32 & \textbf{12.61} & 14.18 \\
Korean      & 13.42 & 13.56 & \textbf{12.12} & 13.76 \\
Spanish     & 42.64 & 41.25 & 41.78 & \textbf{39.44} \\
Vietnamese  & 61.21 & 62.78 & 58.68 & \textbf{53.84} \\
\midrule
Mean & 32.11 & 32.59 & 31.40 & \textbf{28.21} \\
\bottomrule
\end{tabular}
\caption{WER (\%) on L2-Arctic for English speech from speakers with different first-language (L1) backgrounds. Best result per row is in \textbf{bold}.}
\label{tab:wer-l1-speakers}
\end{table}

\section{Experiments}
\label{sec:experiments}
\subsection{Experimental Setup}

We evaluate ASR-TRA on the lightweight Whisper-Tiny model~\cite{radford2023whisper}, which contains approximately 39M parameters and is widely adopted in real-world ASR applications due to its low computational cost. Its compact size makes it well-suited for deployment in latency-sensitive or resource-constrained environments, such as on-device transcription or streaming ASR. However, Whisper-Tiny remains highly sensitive to distribution shifts, struggling to maintain performance under acoustic variations.

To enable fast and targeted adaptation, we insert a learnable decoder prompt of length four, introducing only $4 \times d = 1{,}536$ additional parameters, where $d = 384$ denotes the decoder embedding dimension.
During adaptation, decoder temperatures are randomly sampled from a uniform range $[0.4, 0.6]$, with a total of 4 candidates generated in parallel by the Whisper-Tiny to encourage diverse hypotheses.
Model parameters are updated using a learning rate $\eta_1$ in the range $10^{-6}$ to $10^{-5}$ depending on data complexity, while the prompt parameters are updated with a $100\times$ larger learning rate $\eta_2$, consistent with findings from prompt-tuning literature~\cite{li2021prefix,lester2021prompt}. All experiments are conducted on a single NVIDIA RTX 6000 Ada GPU.

\subsection{Datasets and Evaluation Metrics}

We consider two challenging distribution-shift scenarios that test robustness to both acoustic and linguistic variation:

\textbf{1) Librispeech test-other with Environmental Noise.} 
To evaluate robustness under acoustic corruption, we augment the LibriSpeech test-other set~\cite{panayotov2015librispeech} with background noise sampled from the MS-SNSD corpus~\cite{reddy2019mssnsd}. Eight additive noise types at 10~dB SNR are used: air conditioner, airport announcement, babble, copy machine, munching, neighbors, door shutting, and typing, following prior robustness benchmarks~\cite{hendrycks2019ood,kim2023sgem}. Each utterance is augmented with one randomly sampled instance per noise type to approximate real-world environments.

\textbf{2) L2-Arctic Non-Native Accented Speech.} 
To assess sensitivity to speaker and pronunciation variation, we use the L2-Arctic dataset~\cite{zhao2018l2arctic}, which contains English speech from speakers of six different first-language backgrounds. 
This condition introduces severe accent shifts that Whisper has not encountered during pretraining, thereby posing a strong challenge for evaluating its cross-speaker generalization ability.

These two settings together evaluate the adaptability of our method to acoustic domain shifts, a key challenge for real-world ASR deployment~\cite{wang2021tent,kim2023sgem}.

\paragraph{Evaluation Metrics.}
We report Word Error Rate (WER) for all experiments in this paper, following standard practice in test-time adaptation for ASR~\cite{lin2022suta,kim2023sgem}. WER measures the minimum number of word-level insertions, deletions, and substitutions required to transform the model's transcription into the ground truth, normalized by the number of words in the reference transcript. Lower WER indicates better transcription accuracy.
We also report latency in seconds to quantify the overhead introduced by the proposed TTA process.

\subsection{Comparison to State-of-the-Art Approaches}

We evaluate our method under two major types of distribution shift: environmental noise and speaker accent variation. Results are summarized in Tables~\ref{tab:main-results} and~\ref{tab:wer-l1-speakers}, following prior work on test-time adaptation for ASR~\cite{lin2022suta,kim2023sgem,wang2021tent}.

\paragraph{Noise Robustness.}
We compare our approach with three baselines: (1) the original Whisper-Tiny model without adaptation (baseline), (2) SUTA~\cite{lin2022suta}, and (3) SGEM~\cite{kim2023sgem}. Table~\ref{tab:main-results} reports the word error rate (WER) and inference latency on eight MS-SNSD noise conditions~\cite{reddy2019mssnsd} applied to LibriSpeech test-other~\cite{panayotov2015librispeech}.  
Our method achieves the lowest average WER (28.64\%) and latency (0.720~s), consistently outperforming the baselines. Improvements are particularly pronounced under high-entropy noise such as \textit{airport announcement} and \textit{babble}, where acoustic variability is high. While SUTA marginally outperforms our method under the \textit{neighbors} condition, our method remains significantly faster at inference time. This aligns with findings in reward-guided adaptation literature, where additional diversity helps stabilize learning under heavy perturbations~\cite{lee2025bitta}.

\paragraph{Accent Robustness.}
We further evaluate robustness to speaker and pronunciation variation using the L2-Arctic dataset~\cite{zhao2018l2arctic}, which contains non-native English speakers from six different first-language (L1) backgrounds. Table~\ref{tab:wer-l1-speakers} shows that our method achieves the best mean WER (28.21\%), with substantial improvements on more challenging L1 groups such as Arabic and Vietnamese. These results suggest that our adaptation framework generalizes well across diverse phonetic systems, narrowing the gap between native and non-native speech recognition.

Overall, these findings demonstrate that our method enhances Whisper’s robustness to both acoustic and linguistic distribution shifts while maintaining low inference latency. Compared with entropy-minimization-based methods (SUTA, SGEM), our reward-driven adaptation achieves a better balance between accuracy and efficiency, which is crucial for real-world ASR deployment.

\begin{table*}[t]
\centering
\setlength{\tabcolsep}{5pt}
\begin{tabular}{ccc|cc|cc}
\toprule
\multicolumn{3}{c|}{Configuration} & \multicolumn{2}{c|}{Whisper-Tiny} & \multicolumn{2}{c}{Whisper-Base} \\
Finetune & Prompt & Reward & WER(\%) $\downarrow$ & Latency (s) & WER(\%) $\downarrow$ & Latency (s) \\
\midrule
N & N & N            & 45.06 & --     & 40.25 & --     \\
Y & N & CLAP         & 40.49 & 0.486  & 35.97 & 0.573  \\
N & Y & CLAP         & 42.75 & 0.472  & 37.45 & 0.552  \\
Y & Y & CLAP         & 36.65 & 0.489  & 33.42 & 0.580  \\
Y & Y & LLM          & 36.35 & 4.191  & \textbf{30.24} & 4.355  \\
Y & Y & CLAP + LLM   & \textbf{35.49} & 4.365  & 30.28 & 4.367  \\
\bottomrule
\end{tabular}
\caption{
Ablation study of accuracy--latency trade-offs under different adaptation configurations on noisy LibriSpeech test-other.
Each row enables a combination of prompt tuning, parameter finetuning, and reward feedback (CLAP, LLM, or both).
WER ($\downarrow$) and average inference latency (s/utterance) are reported for Whisper-Tiny and Whisper-Base.
}
\label{tab:ablation-wer-latency}
\end{table*}
\subsection{Ablation Study}

To better understand the contribution of each component, we conduct an ablation study on the LibriSpeech test-other dataset augmented with Gaussian noise at an SNR of 10 dB. In addition to Whisper-Tiny, we include Whisper-Base (74M vs. 39M parameters) to examine the effect of model size. It offers better recognition under noise while retaining efficient inference. We systematically vary three design dimensions in our adaptation framework:
\begin{itemize}
    \item \textbf{Prompt Tuning}: Whether a learnable decoder prompt is injected during decoding.
    \item \textbf{Model Finetuning}: Whether model parameters are updated during test-time adaptation.
    \item \textbf{Reward Modeling}: Whether reward feedback is used, and whether it comes from CLAP (audio-text) or a pretrained LLM (text-text; we use DeepSeek V3~\cite{deepseekv3}), or a combination of both.
\end{itemize}
As shown in Table~\ref{tab:ablation-wer-latency}, all adaptation configurations outperform the unadapted Whisper baseline, highlighting the effectiveness of both prompt tuning and reward feedback. Several key observations emerge:

\paragraph{Effectiveness of Prompt Tuning.}
Comparing the second row (finetuning only, CLAP reward) and the fourth row (finetuning + prompt tuning, CLAP reward), we observe a notable WER reduction (40.49\%$\rightarrow$36.65\% for Whisper-Tiny) accompanied by only marginal increases in inference latency. This suggests that the decoder prompt provides complementary adaptation capacity to parameter updates, improving alignment with reward signals.

\paragraph{Reward Modeling.}
Using CLAP as a reward model enables substantial WER improvements at negligible latency cost. Incorporating LLM-based feedback yields further gains (e.g., 0.3342$\rightarrow$0.3024 for Whisper-Base) but adds a 7--9$\times$ latency overhead. The hybrid CLAP+LLM reward achieves the best overall WER. We also find that CLAP scores are negatively correlated with ground-truth WER (Spearman $\rho=-0.431$), supporting its reliability as a semantic reward.

\paragraph{Efficiency Considerations.}
We also explored LoRA-based parameter-efficient adaptation~\cite{hu2021lora} but observed negligible speedup compared to full finetuning, as the majority of runtime is dominated by autoregressive decoding and reward computation. We therefore omit LoRA from further comparisons and instead focus on balancing reward complexity with inference efficiency.

\subsection{Subset Evaluation on Confident Samples}

We further analyze a subset of the LibriSpeech test-other set~\cite{panayotov2015librispeech}, comprising the 100 samples with the highest model confidence under additive Gaussian noise. Interestingly, the baseline WER on these high-confidence samples is as high as \textbf{83.61\%}, which is \textit{worse} than the WER on the full test set. This counterintuitive result indicates that Whisper-Tiny is often highly confident in its incorrect predictions, a phenomenon we refer to as \textit{blind confidence}, where the model’s internal certainty fails to reflect actual transcription accuracy under distribution shift.

As shown in Figure~\ref{fig:confident-subset}, SUTA~\cite{lin2022suta}, which strongly depends on confidence-based entropy minimization, dramatically degrades performance to \textbf{122.37\%}, exacerbating the misalignment between confidence and accuracy. SGEM~\cite{kim2023sgem}, which incorporates sequence-level uncertainty, performs better at \textbf{63.00\%}, but still fails to fully address blind confidence.

Our method ASR-TRA achieves the lowest WER of \textbf{45.17\%} on this challenging subset, reducing errors by nearly half compared to the baseline. This robustness stems from the fact that our adaptation does not rely on model confidence or entropy as internal signals. Instead, it leverages external reward models (e.g., CLAP and LLM) to evaluate transcription quality, enabling it to revise predictions even when the model is falsely confident.
By decoupling adaptation decisions from the model’s own uncertainty estimates, our approach mitigates the effects of blind confidence.

These results reveal that internal confidence measures in Whisper-Tiny are not reliable under distribution shift, and that methods relying on confidence may be brittle when confronted with misleading inputs. Reward-guided adaptation provides a more reliable criterion, decoupling the adaptation signal from the model’s own uncertainty estimates.

\section{Discussion}

Our experiments show that integrating causal reasoning with reinforcement learning enables effective test-time adaptation for Whisper-based ASR. Unlike confidence-driven baselines such as entropy minimization~\cite{wang2021tent,lin2022suta} and pseudo-labeling~\cite{lee2013pseudo,sohn2020fixmatch}, our causal intervention framework avoids error amplification from unreliable confidence signals and provides clearer interpretability, as confirmed by confidence-subset analysis where methods like SUTA tended to reinforce mistakes. Moreover, incorporating complementary external rewards further improves robustness to acoustic and linguistic shifts: CLAP offers fast and stable audio guidance, while DeepSeek V3 provides more precise semantic feedback at higher computational cost. Together, these components highlight the benefit of leveraging external cues rather than relying solely on internal confidence estimates for reliable adaptation under distribution shift.

Two limitations remain. First, our reward models have inherent constraints: LLM-based rewards dominate inference latency in hybrid settings, while CLAP currently supports mainly English audio–text similarity, limiting multilingual evaluation. Second, the framework focuses on single-utterance adaptation; extending to streaming or conversational ASR could enable persistent, context-aware robustness in real-world applications. Such a setting also offers the potential for implicit few-shot learning, where temporally accumulated feedback across utterances provides supervision akin to few-shot prompting.

\section{Conclusion}

We presented a causal reinforcement-learning framework, ASR-TRA, that adapts Whisper during inference through three lightweight stages. The adaptation begins by injecting a learnable decoder prompt as an explicit causal intervention on the decoding trajectory. To explore alternatives, the model generates diverse transcriptions via temperature-controlled sampling, which are then scored by CLAP to provide rewards. These rewards drive a policy-gradient update that fine-tunes the prompt and model parameters, gradually steering the model toward semantically improved outputs.

This loop consistently reduces WER on noisy and accented speech while introducing limited additional latency compared with existing test-time adaptation methods. By avoiding reliance on model-internal confidence and instead using an external, modality-aligned reward signal, the approach remains interpretable and robust under distribution shift. More broadly, framing test-time adaptation as a reward-driven causal process provides a promising direction for practical on-device or low-resource ASR, and suggests a natural path toward tighter integration of speech recognition with downstream multimodal or conversational systems.

\section*{Acknowledgments}
This work was supported by the National Natural Science Foundation of China (No. 62471420), GuangDong Basic and Applied Basic Research Foundation (2025A1515012296), and 2025 Tencent AI Lab Rhino-Bird Program.

\bibliography{aaai2026}

\end{document}